# Cell Press Multi-Journal Submission

## Global, Regional, and National Burden of Chronic Kidney Disease Attributable to High Body Mass Index among Individuals Aged 20-54 Years from 1990 to 2021: An Analysis of the Global Burden of Disease Study

--Manuscript Draft--

| | |
|---|---|
| **Manuscript Number:** | |
| **Full Title:** | Global, Regional, and National Burden of Chronic Kidney Disease Attributable to High Body Mass Index among Individuals Aged 20-54 Years from 1990 to 2021: An Analysis of the Global Burden of Disease Study |
| **Article Type:** | Research Article |
| **Keywords:** | chronic kidney disease; high body mass index; global burden; mortality; disability adjusted life year |
| **Corresponding Author:** | Ying Zhang<br>Shanghai Sixth People's Hospital<br>Bethesda, CHINA |
| **Order of Authors:** | Guangxi Wu |
| | Yu Chen |
| | Wei Ma |
| | Ying Zhang |
| **Abstract:** | Background<br><br>Chronic kidney disease (CKD) is one of the most prevalent non-communicable health issues globally, and high body mass index (BMI) plays a significant role in the onset and progression of CKD. To date, there has been limited attention given to the global burden of CKD attributable to high BMI. This study aims to report the global, regional, and national burden of CKD-related deaths and disability-adjusted life years (DALYs) attributable to high BMI among individuals aged 20-54 years from 1990 to 2021.<br><br>Methods<br><br>Data on the disease burden attributable to high BMI were retrieved from the 2021 Global Burden of Disease, Injuries, and Risk Factors Study (GBD). The global cases, age-standardized mortality rate (ASMR), and age-standardized disability-adjusted life years (ASDR) attributable to high BMI were estimated based on age, sex, geographic location, and the Social-demographic Index (SDI). The estimated annual percentage change (EAPC) was calculated to quantify trends in ASMR and ASDR from 1990 to 2021. Decomposition and frontier analyses were conducted to understand the drivers behind changes in burden and to identify top-performing countries. Inequality analysis was performed to assess disparities in burden across different SDI levels. The Bayesian age-period-cohort (BAPC) model was used to predict the disease burden up to 2035.<br><br>Results<br><br>In 2021, the global deaths and DALYs attributable to high BMI-related CKD are more than triple the figures from 1990. Additionally, from 1990 to 2021, the ASMR and ASDR accelerated respectively, particularly among males, in High-income North America and in Low-middle SDI regions. In terms of SDI, the Low-middle SDI region had the highest ASMR and ASDR related to CKD in 2021. It is projected that similar patterns will persist over the next decade without intervention. Due to higher obesity rates, population growth, and insufficient healthcare resources, Low-middle SDI regions face higher EAPCs. Furthermore, this inequality has become more pronounced. Males across all age groups exhibited higher mortality rates.<br><br>Conclusion |

|  | From 1990 to 2021, there was a significant increase in global deaths and DALYs attributable to high BMI-related CKD. As a major public health issue for CKD patients, high BMI urgently requires targeted measures to address it. |
|---|---|
| **Opposed Reviewers:** |  |
| **Suggested Reviewers:** |  |
| **Additional Information:** |  |
| **Question** | **Response** |
| **Standardized datasets** <br> A list of datatypes considered standardized under Cell Press policy is available here. Does this manuscript report new standardized datasets? | No |
| **Original code** <br> Does this manuscript report original computer code, algorithms, or computational models? | No |

Dear Editor,:

I am writing to submit our manuscript entitled, "Global, Regional, and National Burden of Chronic Kidney Disease Attributable to High Body Mass Index among Individuals Aged 20-54 Years from 1990 to 2021: An Analysis of the Global Burden of Disease Study" for consideration as a public health research article for Cell Press Multi-Journal. This study examines the impact of high body mass index on the burden of chronic kidney disease at multiple geographic levels over three decades, highlighting patterns and trends in mortality and disability.

Given the rising prevalence of obesity and its contribution to non-communicable diseases, we believe that our findings will be of interest to the broad readership of Cell Press Multi-Journal's submission, who seek to understand complex health challenges worldwide. Our study provides valuable insights into the demographic and regional disparities in chronic kidney disease burden related to high BMI, emphasizing the urgent need for targeted interventions to mitigate this public health concern.

Each of the authors confirms that this manuscript has not been previously published and is not currently under consideration by any other journal. Additionally, all of the authors have approved the contents of this paper and have agreed to the Cell Press Multi-Journal's submission policies.

Each named author has substantially contributed to conducting the underlying research and drafting this manuscript. Additionally, to the best of our knowledge, the named authors have no conflict of interest, financial or otherwise.

Sincerely,

Guangxi Wu

International Peace Maternity and Child Health Hospital, Shanghai Jiao Tong University School of Medicine

Hengshan Road 910, Shanghai, China.

wuguangxi0713@163.com

# Global, Regional, and National Burden of Chronic Kidney Disease Attributable to High Body Mass Index among Individuals Aged 20-54 Years from 1990 to 2021: An Analysis of the Global Burden of Disease Study

Yu Chen[1,#], Guangxi Wu[2,3,#], Wei Ma[2,3,*], Ying Zhang[4,*]

1, Department of Anesthesiology, The Third People's Hospital of Bengbu, Bengbu Medical University, Bengbu, China.
2, Department of Anesthesiology, International Peace Maternity and Child Health Hospital, Shanghai Jiao Tong University School of Medicine, Shanghai, China.
3, Shanghai Key Laboratory of Embryo Original Diseases, Shanghai, China.
4, Department of Anesthesiology, Shanghai Sixth People's Hospital Affiliated to Shanghai Jiao Tong University School of Medicine

#Equal contribution
Yu Chen and Guangxi Wu

*Co-Corresponding authors:
Wei Ma
Department of Anesthesiology, International Peace Maternity and Child Health Hospital, Shanghai Jiao Tong University School of Medicine, Shanghai, China.
Email: shemar_sh@126.com

Ying Zhang
Department of Anesthesiology, Shanghai Sixth People's Hospital Affiliated to Shanghai Jiao Tong University School of Medicine, Shanghai, China.
Email: zhang198069ying@163.com

Other author's email:
Yu Chen, chenyu19880222@163.com
Guangxi Wu, wuguangxi0713@163.com

**ABSTRACT**

**Background**: Chronic kidney disease (CKD) is one of the most prevalent non-communicable health issues globally, and high body mass index (BMI) plays a significant role in the onset and progression of CKD. To date, there has been limited attention given to the global burden of CKD attributable to high BMI. This study aims to report the global, regional, and national burden of CKD-related deaths and disability-adjusted life years (DALYs) attributable to high BMI among individuals aged 20-54 years from 1990 to 2021.

**Methods**: Data on the disease burden attributable to high BMI were retrieved from the 2021 Global Burden of Disease, Injuries, and Risk Factors Study (GBD). The global cases, age-standardized mortality rate (ASMR), and age-standardized disability-adjusted life years (ASDR) attributable to high BMI were estimated based on age, sex, geographic location, and the Social-demographic Index (SDI). The estimated annual percentage change (EAPC) was calculated to quantify trends in ASMR and ASDR from 1990 to 2021. Decomposition and frontier analyses were conducted to understand the drivers behind changes in burden and to identify top-performing countries. Inequality analysis was performed to assess disparities in burden across different SDI levels. The Bayesian age-period-cohort (BAPC) model was used to predict the disease burden up to 2035.

**Results**: In 2021, the global deaths and DALYs attributable to high BMI-related CKD are more than triple the figures from 1990. Additionally, from 1990 to 2021, the ASMR and ASDR accelerated respectively, particularly among males, in High-income North America and in Low-middle SDI regions. In terms of SDI, the Low-middle SDI region had the highest ASMR and ASDR related to CKD in 2021. It is projected that similar patterns will persist over the next decade without intervention. Due to higher obesity rates, population growth, and insufficient healthcare resources, Low-middle SDI regions face higher EAPCs. Furthermore, this inequality has become more pronounced. Males across all age groups exhibited higher mortality rates.

Conclusion: From 1990 to 2021, there was a significant increase in global deaths and DALYs attributable to high BMI-related CKD. As a major public health issue for CKD patients, high BMI urgently requires targeted measures to address it.

## 1.INTRODUCTION

Chronic Kidney Disease (CKD) is a significant contributor to the morbidity and mortality of non-communicable diseases, imposing a heavy burden on public health worldwide [1]. In 2019, CKD was the 11th leading cause of death globally, with over 1.43 million deaths. In the worst-case scenario, this number is projected to increase to 4 million by 2040 [2, 3]. CKD is a major cause of disease and economic burden due to the high cost of renal replacement therapy for end-stage kidney disease (ESKD) [4]. Patients with CKD or ESKD face challenges such as reduced quality of life, increased healthcare costs, and greater economic burdens[5, 6] .

Obesity has become a serious public health issue. In 2015, obesity affected

603.7 million adults globally [7]. Furthermore, from 1990 to 2017, there was a significant increase in BMI-related mortality and disability-adjusted life years (DALYs) worldwide [8]. Similarly, in the United States alone, the disease burden associated with high BMI is estimated to cause 216,000 deaths annually, with a cost of $113.9 billion [9]. Obesity is a modifiable metabolic risk factor characterized by a long-term positive energy balance leading to excessive fat accumulation. The etiology of obesity is multifactorial, involving genetic, environmental, socioeconomic, and behavioral factors. These factors contribute to low-grade chronic inflammation, abnormal hormonal and immune responses, and ultimately systemic metabolic dysregulation [10-12]. Obesity significantly increases the risk of developing CKD [13]. Treating obesity is challenging because patients often find it difficult to adhere to strict diet and exercise regimens. Pharmacological treatments and bariatric surgery for obesity are expensive and may lead to various adverse effects such as nausea, vomiting, diarrhea, and neuropathy[14].

The Global Burden of Disease (GBD) 2021 Study provides a comprehensive dataset to assess the burden of diseases and injuries, including those related to CKD. This analysis aims to quantify the global, regional, and national burden of mortality and DALYs caused by CKD, with a focus on high BMI as a risk factor. By analyzing temporal trends in disease burden, we can identify regions experiencing the greatest increases, thereby informing disease control and prevention strategies.

Therefore, this study aims to describe the global trends in CKD incidence and DALYs attributable to high BMI from 1990 to 2021, stratifying global trends by age group, sex, and Socio-demographic Index (SDI), and revealing regional and national trends. This study will also explore the correlation between CKD burden and SDI socioeconomic indicators to assess the impact of socioeconomic factors on health outcomes.

Additionally, this study aims to evaluate the contributions of aging, population, and epidemiological factors to CKD attributable to high BMI, analyze inequalities among countries, and project changes up to 2035. The findings of this analysis will provide critical information for public health policies and initiatives aimed at reducing the burden of CKD associated with high BMI.

## 2.MATERIALS AND METHODS
**Data collection**

The data for this investigation were sourced from the Global Health Data Exchange (GBD) Results Tool (http://ghdx.healthdata.org/gbd-results-tool), which compiles information on the incidence, prevalence, mortality, and disability-adjusted life years (DALYs) for 369 diseases and injuries, as well as comparative risk assessments for 87 risk factors, across 204 countries and regions from 1990 to 2021. The 2021 global biodiversity data were collected through systematic evaluations of population censuses, household surveys, civil registration and vital statistics, disease registries, disease notifications, healthcare utilization, air pollution monitoring, and satellite imagery [15].

This study includes data on CKD (Chronic Kidney Disease) deaths, total DALYs, crude mortality rates, crude DALY rates, age-standardized mortality rates (ASMR) for T2CKD, DALYs per 100,000 population attributable to high body mass index (BMI), and the Socio-demographic Index (SDI) across 204 countries and regions from 1990 to 2021. DALYs represent the sum of years of life lost due to premature death and years lived with disability, with one DALY equivalent to one lost year of healthy life[16]. The SDI, introduced in the 2015 World Population Development Report, is an index reflecting the developmental status of a geographic location. It is calculated based on the lag-distributed income per capita, the average education level of individuals aged 15 and above, and the total fertility rate for those under 25 years old [17]. Based on SDI quintiles, the 204 countries and regions were categorized into five groups: high SDI (>0.805129), high-middle SDI (0.689504–0.805129), middle SDI (0.607679–0.689504), low-middle SDI (0.454743–0.607679), and low SDI (≤0.454743) [18].

**Study population**

The study population (20-54 years old) includes both young and middle-aged individuals. The data were stratified by age groups (20-24, 25-29, 30-34, 35-39, 40-44, 45-49, and 50-54 years).

Definitions of chronic kidney disease and high BMI

Chronic Kidney Disease (CKD) was defined based on an estimated

glomerular filtration rate (eGFR) of < 60 mL/min/1.73 m² and/or a urine albumin-to-creatinine ratio (ACR) ≥ 30 mg/g[19] (all stages, including kidney replacement therapy (KRT), as detailed in Supplementary Data, Table S1).

Meanwhile, high BMI was defined as a BMI > 25 kg/m² for adults (aged 20 years and older), and for individuals aged 1–19 years, thresholds based on the International Obesity Task Force standards were used[20, 21].

**Statistical analysis**

The Estimated Annual Percentage Change (EAPC) is widely used to measure trends in age-standardized rates over a specific period [22]. This method was employed to assess the trends in ASMR (Age-Standardized Mortality Rate) and ASDR (Age-Standardized DALY Rate) attributable to high BMI from 1990 to 2021. The calculation was performed using a linear regression model as follows [23]:

$$\ln(ASDR or ASMR) = \alpha + \beta x + \varepsilon$$

$$EAPC = 100 \times (\exp(\beta) - 1)$$

In the above equation, x represents the calendar year, and ε is the error term. β describes the positive or negative trend in age-standardized rates. Based on this model, the Estimated Annual Percentage Change (EAPC) and its 95% confidence interval (CI) can be derived. If both the EAPC value and its 95% CI are greater than 0, the trend is considered to be increasing. If both the EAPC value and its 95% CI are less than 0, the trend is considered to be decreasing.

Otherwise, the burden of CKD attributable to high BMI is considered stable.

Additionally, the number of CKD deaths or DALYs, ASMR or ASDR, and percentage changes with 95% uncertainty intervals (UIs) were estimated by sex, age, location, and SDI to comprehensively assess the burden of CKD attributable to high BMI. The GBD database quantified these estimates over 1,000 iterations using the Bayesian meta-regression tool DisMod-MR 2.1[8], with 95% UIs defined by the 25th and 975th values of the ordered 1,000 iterations. Furthermore, Spearman's rank correlation test was conducted to evaluate the relationship between high BMI-related CKD burden (ASMR and ASDR) and the 2021 SDI. Additionally, the Das Gupta decomposition method [23] was employed to decompose global CKD deaths and DALYs based on age structure, population growth, and epidemiological changes. All statistical analyses and graphical visualizations were performed using R software version 4.4.2.

## 3.RESULTS

### Global CKD burden attributable to high BMI from 1990 to 2021

In 2021, the global number of CKD deaths attributable to high BMI was 44,643.41, with an ASMR of 2.91 per 100,000 population, representing a 271.4% increase in deaths since 1990 (Table 1). Additionally, the number of DALYs increased by 248.3% over the past 30 years, with an ASDR of 122.08 per 100,000 population in 2021 (Table 2). From 1990 to 2021, the EAPC for ASMR and ASDR were 2.25 and 1.98, respectively, indicating a worsening burden of

CKD due to high BMI.

Among the 204 countries and regions, the highest ASMR and ASDR for CKD attributable to high BMI in 2021 occurred in Saudi Arabia (35.62 and 724.26 per 100,000 population, respectively), while Ukraine had the lowest ASMR (0.56 per 100,000 population) and Finland had the lowest ASDR (39.78 per 100,000 population) (Figure 1A, 2; Supplementary Tables S2, S3). Between 1990 and 2021, Ukraine experienced the largest increase in ASMR for CKD burden related to high BMI (2202.41%), while Poland saw the largest decrease (-40.14%) (Supplementary Table S2).

Furthermore, Lesotho had the largest increase in ASDR (255.37%), followed by El Salvador (223.85%), while Poland had the largest decrease (-35.86%) (Supplementary Table S3). During the period from 1990 to 2021, the three countries with the largest declines in CKD deaths attributable to high BMI were the Philippines (-54.75%), Cyprus (-51.1%), and Georgia (-42.45%). Between 1990 and 2021, the number of DALYs increased in the vast majority of the 204 countries and regions, with the United Arab Emirates experiencing the largest increase (1,043.5%) (Figure 1B).

Additionally, a significant negative correlation was observed between ASMR (r = -0.1677, p < 0.00001) and ASDR (r = -0.2107, p < 0.00001) attributable to high BMI and SDI in 2021 (Figure 2).

TABLE 1 Deaths and ASMR of CKD attributable to high BMI and the temporal trends from 1990 to 2021.

## Table 1: Deaths and ASMR of CKD attributable to high BMI and the temporal trends from 1990 to 2021

| location | 1990 deaths cases no. (95% UI) | 1990 ASMR per 100,000 no. (95% UI) | 2021 deaths cases no. (95% UI) | 2021 ASMR per 100,000 no. (95% UI) | 1990-2021 EAPC in ASMR no. (95% CI) |
|---|---|---|---|---|---|
| Global | 12021.18 (6113.05, 18620.94) | 44643.41 (24459.85, 65047.52) | 44643.41 (24459.85, 65047.52) | 5.06 (2.70, 7.51) | 2.25 (2.13, 2.37) |
| Andean Latin America | 197.31 (104.24, 295.08) | 776.70 (477.67, 1121.53) | 776.70 (477.67, 1121.53) | 14.72 (8.23, 21.55) | 1.88 (1.52, 2.25) |
| Australasia | 16.29 (7.74, 27.49) | 49.38 (25.01, 76.88) | 49.38 (25.01, 76.88) | 3.64 (1.98, 5.35) | 2.07 (1.80, 2.35) |
| Caribbean | 207.83 (114.09, 289.67) | 699.12 (418.14, 962.79) | 699.12 (418.14, 962.79) | 9.33 (5.17, 13.92) | 2.68 (2.48, 2.87) |
| Central Asia | 127.95 (60.54, 214.37) | 540.22 (256.84, 911.08) | 540.22 (256.84, 911.08) | 3.14 (1.58, 5.02) | 3.31 (2.80, 3.83) |
| Central Europe | 439.53 (208.14, 725.88) | 336.84 (168.86, 527.47) | 336.84 (168.86, 527.47) | 3.30 (1.82, 4.91) | 0.40 (0.20, 0.59) |
| Central Latin America | 912.75 (482.28, 1397.46) | 5600.97 (3200.13, 8328.63) | 5600.97 (3200.13, 8328.63) | 15.01 (8.52, 21.75) | 2.78 (2.23, 3.34) |
| Central Sub-Saharan Africa | 200.47 (92.14, 342.08) | 985.44 (462.21, 1684.38) | 985.44 (462.21, 1684.38) | 9.19 (4.50, 16.02) | 1.57 (1.43, 1.70) |
| East Asia | 1901.62 (837.53, 3651.48) | 4302.14 (2055.98, 7225.98) | 4302.14 (2055.98, 7225.98) | 2.91 (1.44, 4.88) | 1.38 (1.27, 1.49) |
| Eastern Europe | 424.46 (182.46, 740.95) | 418.43 (189.26, 732.27) | 418.43 (189.26, 732.27) | 1.57 (0.84, 2.33) | 2.65 (2.26, 3.05) |
| Eastern Sub-Saharan Africa | 312.56 (134.02, 562.27) | 1321.91 (602.30, 2320.51) | 1321.91 (602.30, 2320.51) | 5.22 (2.44, 8.76) | 1.56 (1.49, 1.64) |
| High SDI | 1557.75 (857.00, 2255.35) | 5888.53 (3628.79, 7750.82) | 5888.53 (3628.79, 7750.82) | 5.06 (2.75, 7.39) | 2.75 (2.59, 2.91) |
| High-income Asia Pacific | 209.31 (102.33, 349.86) | 162.56 (80.00, 264.96) | 162.56 (80.00, 264.96) | 1.97 (0.93, 3.19) | −0.20 (−0.32, −0.09) |
| High-income North America | 659.10 (379.87, 866.70) | 3256.37 (2112.01, 4040.97) | 3256.37 (2112.01, 4040.97) | 9.26 (5.07, 13.12) | 4.09 (3.86, 4.31) |
| High-middle SDI | 2224.61 (1119.64, 3616.11) | 4506.34 (2458.45, 6653.90) | 4506.34 (2458.45, 6653.90) | 3.84 (2.06, 5.76) | 1.48 (1.39, 1.57) |
| Low SDI | 1059.94 (504.85, 1775.07) | 4281.07 (2086.79, 6946.54) | 4281.07 (2086.79, 6946.54) | 4.31 (2.09, 6.98) | 1.30 (1.17, 1.42) |
| Low-middle SDI | 2415.68 (1206.70, 3835.67) | 11857.15 (6261.03, 18074.89) | 11857.15 (6261.03, 18074.89) | 5.31 (2.78, 8.04) | 2.46 (2.39, 2.53) |
| Middle SDI | 4744.86 (2361.82, 7454.90) | 18063.01 (9934.06, 26475.25) | 18063.01 (9934.06, 26475.25) | 5.49 (2.92, 8.35) | 2.08 (1.89, 2.27) |
| North Africa and Middle East | 1683.36 (959.08, 2514.76) | 7634.26 (4401.44, 11028.59) | 7634.26 (4401.44, 11028.59) | 14.65 (7.98, 21.36) | 1.86 (1.70, 2.02) |
| Oceania | 21.04 (9.42, 37.52) | 71.93 (30.46, 127.29) | 71.93 (30.46, 127.29) | 5.58 (2.51, 9.51) | 1.35 (1.21, 1.49) |
| South Asia | 1398.88 (606.25, 2392.55) | 7077.04 (3432.05, 11918.32) | 7077.04 (3432.05, 11918.32) | 2.36 (1.20, 3.93) | 2.55 (2.46, 2.63) |
| Southeast Asia | 1061.67 (450.02, 1851.48) | 4941.51 (2140.26, 8588.69) | 4941.51 (2140.26, 8588.69) | 4.11 (1.82, 6.87) | 2.66 (2.57, 2.76) |
| Southern Latin America | 224.69 (117.82, 342.60) | 366.48 (211.51, 532.04) | 366.48 (211.51, 532.04) | 9.08 (5.03, 13.35) | 0.73 (0.35, 1.12) |
| Southern Sub-Saharan Africa | 244.93 (117.64, 398.38) | 1066.08 (566.38, 1647.44) | 1066.08 (566.38, 1647.44) | 11.12 (5.60, 17.17) | 3.25 (2.78, 3.72) |
| Tropical Latin America | 924.23 (506.99, 1312.24) | 1976.90 (1220.04, 2666.01) | 1976.90 (1220.04, 2666.01) | 8.12 (4.61, 11.22) | 1.28 (1.08, 1.48) |
| Western Europe | 262.47 (116.63, 439.16) | 281.23 (138.04, 458.71) | 281.23 (138.04, 458.71) | 3.23 (1.56, 4.98) | 1.85 (1.67, 2.04) |
| Western Sub-Saharan Africa | 590.58 (269.67, 989.59) | 2777.90 (1361.80, 4615.66) | 2777.90 (1361.80, 4615.66) | 8.17 (3.94, 12.90) | 1.66 (1.58, 1.73) |

No., number; ASMR, age–standardized mortality rate; UI, uncertainty interval; EAPC, estimated annual percentage change; CI, confidential interval.

TABLE 2 DALYs and ASDR of CKD attributable to high BMI and the temporal trends from 1990 to 2021

| location | 1990 DALYs no. (95% UI) | 1990 ASDR per 100,000 no. (95% UI) | 2021 DALYs no. (95% UI) | 2021 ASDR per 100,000 no. (95% UI) | 1990-2021 EAPC in ASDR no. (95% CI) |
|---|---|---|---|---|---|
| Global | 721827.11 (370977.87, 1119447.65) | 69.13 (35.06, 106.00) | 2514227.16 (1357172.58, 3724507.59) | 122.08 (66.25, 180.18) | 1.98 (1.89, 2.08) |
| Andean Latin America | 10043.97 (5393.66, 15005.51) | 186.21 (97.63, 279.02) | 38130.42 (23230.05, 55379.70) | 391.24 (227.14, 565.48) | 2.78 (2.26, 3.31) |
| Australasia | 1546.92 (814.21, 2476.98) | 53.97 (28.18, 81.36) | 4217.18 (2381.71, 6401.69) | 76.16 (43.49, 106.96) | 1.48 (1.29, 1.67) |
| Caribbean | 11300.41 (6263.94, 16131.55) | 126.29 (69.09, 183.86) | 35594.77 (21430.45, 48869.23) | 234.81 (135.40, 339.31) | 2.55 (2.39, 2.72) |
| Central Asia | 12229.01 (6278.55, 19430.47) | 68.68 (37.39, 102.79) | 40929.19 (20817.04, 65390.04) | 127.70 (67.69, 193.21) | 1.84 (1.56, 2.13) |
| Central Europe | 28124.81 (13758.71, 44241.40) | 89.00 (48.98, 131.40) | 25397.16 (13884.40, 38217.09) | 89.32 (50.24, 130.68) | 0.17 (0.04, 0.29) |
| Central Latin America | 54957.41 (29155.30, 84568.67) | 186.82 (99.35, 269.38) | 288959.45 (165687.44, 424425.26) | 320.22 (188.47, 454.36) | 1.71 (1.37, 2.04) |
| Central Sub-Saharan Africa | 10555.60 (4974.55, 17888.09) | 115.81 (59.71, 183.72) | 51517.17 (24654.47, 87171.38) | 268.33 (140.76, 408.08) | 2.73 (2.34, 3.12) |
| East Asia | 102590.24 (45149.35, 199259.13) | 44.84 (20.23, 85.00) | 249712.31 (125134.93, 406449.63) | 68.24 (34.25, 112.21) | 1.42 (1.29, 1.55) |
| Eastern Europe | 31162.18 (15340.58, 51122.16) | 46.60 (22.80, 76.07) | 31340.48 (16058.92, 51254.17) | 43.85 (21.26, 70.04) | −0.16 (−0.25, −0.07) |
| Eastern Sub-Saharan Africa | 15709.05 (6805.07, 27623.11) | 72.37 (33.49, 127.01) | 68041.50 (31129.76, 117381.61) | 117.81 (56.44, 195.00) | 1.43 (1.36, 1.50) |
| High SDI | 121928.89 (66144.63, 180362.89) | 64.68 (33.97, 94.57) | 361295.22 (223909.99, 487952.39) | 119.83 (68.69, 167.37) | 2.26 (2.14, 2.38) |
| High-income Asia Pacific | 15088.45 (7281.42, 24825.65) | 38.91 (20.31, 58.48) | 14458.13 (7078.88, 23338.99) | 57.20 (32.55, 82.88) | 0.95 (0.81, 1.10) |
| High-income North America | 52170.17 (30846.75, 73055.12) | 81.26 (44.79, 112.20) | 191498.31 (123799.10, 245212.68) | 209.80 (123.68, 284.46) | 3.40 (3.19, 3.61) |
| High-middle SDI | 139952.08 (70597.52, 219129.12) | 65.11 (33.26, 99.70) | 281456.75 (153054.05, 419179.17) | 89.04 (47.48, 132.50) | 1.09 (1.02, 1.16) |
| Low SDI | 55952.49 (26160.78, 93442.82) | 71.68 (35.64, 119.85) | 232852.58 (113846.05, 377387.81) | 107.94 (52.88, 174.77) | 1.24 (1.14, 1.34) |
| Low-middle SDI | 135360.40 (68046.18, 215924.57) | 65.55 (33.75, 104.30) | 652514.60 (344494.37, 1011710.48) | 136.22 (71.43, 204.33) | 2.50 (2.42, 2.58) |
| Middle SDI | 267560.61 (136889.11, 422355.46) | 74.99 (37.30, 120.40) | 983553.76 (539860.27, 1462645.37) | 135.25 (72.54, 202.01) | 2.03 (1.83, 2.22) |
| North Africa and Middle East | 88266.16 (50615.66, 133576.67) | 190.68 (104.68, 295.37) | 394205.44 (223829.32, 569381.19) | 311.83 (178.63, 449.15) | 1.72 (1.63, 1.81) |
| Oceania | 1115.56 (501.99, 1930.18) | 95.79 (43.85, 164.39) | 3771.68 (1641.56, 6370.29) | 140.73 (62.80, 233.05) | 1.17 (1.02, 1.31) |
| South Asia | 87189.00 (38538.16, 152421.25) | 148.61 (81.45, 214.84) | 443010.74 (212747.21, 725593.12) | 197.79 (113.71, 269.46) | 0.83 (0.64, 1.03) |
| Southeast Asia | 53922.89 (23289.35, 92085.36) | 50.45 (22.70, 86.09) | 242498.83 (107034.10, 415382.43) | 106.92 (48.70, 176.80) | 2.54 (2.44, 2.64) |
| Southern Latin America | 12161.16 (6538.05, 18491.26) | 165.38 (90.15, 235.43) | 21704.58 (12477.47, 31755.27) | 183.14 (103.23, 260.25) | 0.58 (0.24, 0.91) |
| Southern Sub-Saharan Africa | 14518.13 (7207.26, 23367.34) | 132.20 (66.32, 219.03) | 57237.28 (30510.40, 88092.59) | 218.67 (108.96, 371.98) | 1.45 (1.33, 1.56) |
| Tropical Latin America | 51962.38 (28882.59, 75528.26) | 33.15 (16.14, 55.66) | 112495.94 (68887.97, 152280.75) | 72.97 (36.46, 118.14) | 2.65 (2.59, 2.72) |
| Western Europe | 34931.70 (16752.31, 56826.36) | 55.17 (27.82, 82.99) | 44888.68 (23124.78, 71459.64) | 66.64 (33.40, 99.21) | 0.84 (0.75, 0.93) |
| Western Sub-Saharan Africa | 32281.92 (14880.29, 54138.91) | 111.99 (54.70, 181.56) | 154617.94 (75177.60, 253738.27) | 188.72 (94.72, 297.39) | 1.56 (1.48, 1.64) |

DALYs, disability–adjusted life–years; No., number; ASDR, age–standardized DALY rate; UI, uncertainty interval; EAPC, estimated annual percentage change; CI, confidential interval.

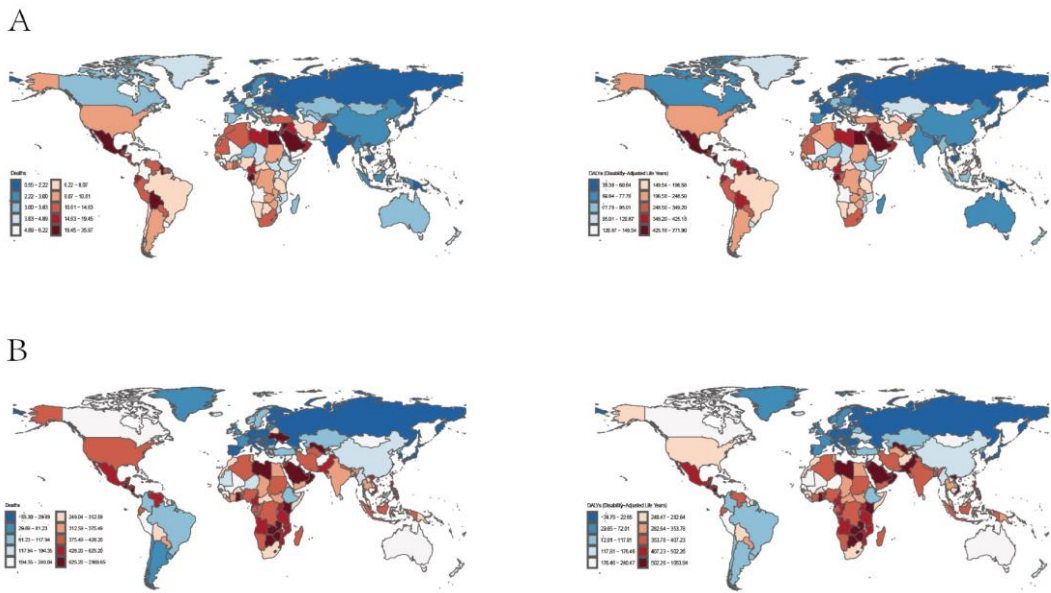

Figure .1. Global CKD burden attributable to high BMI in 204 countries and territories.(A) Age-standardized mortality rate (ASMR) and age-standardized disability-adjusted life year rate (ASDR) in 2021; and (B) The number of deaths and DALYs changed between 1990 and 2021.

**CKD burden attributable to high BMI by 21 GBD regions**

From 1990 to 2021, the ASMR and ASDR for CKD attributable to high BMI increased with rising SDI, peaking when the SDI reached approximately 0.6, then declining, and showing a slight increase again when the SDI reached around 0.8. Among the 21 GBD regions, the highest ASMR and ASDR related to BMI in 2021 were observed in Central Latin America (15.0 and 408.93 per 100,000 population), followed by Andean Latin America (14.7 and 320.2 per 100,000 population) and North Africa and the Middle East (14.6 and 311.8 per 100,000 population) (Tables 1, 2; Figure 3). In contrast, the lowest ASMR and ASDR were found in Eastern Europe (1.57 and 43.85 per 100,000 population), followed by High-income Asia Pacific (1.97 and 57.2 per 100,000 population)

(Tables 1, 2; Figure 3). From 1990 to 2021, the EAPC for ASMR in High-income Asia Pacific was -0.20, indicating a slight decline in CKD ASMR attributable to high BMI in this region over the past 30 years. Similarly, the EAPC for ASDR in Eastern Europe was -0.16, reflecting a slight decrease over the same period. In contrast, all other 21 GBD regions experienced varying degrees of increase in ASDR.

The annual time series from 1990 to 2021 illustrates the observed global and regional age-standardized mortality rates (ASMR) and DALY rates in relation to the Socio-demographic Index (SDI) (Figure 3). Most regions generally followed an upward trend in mortality and DALY rates. However, during the study period, the vast majority of regions experienced an increasing trend in age-standardized mortality rates and DALYs.

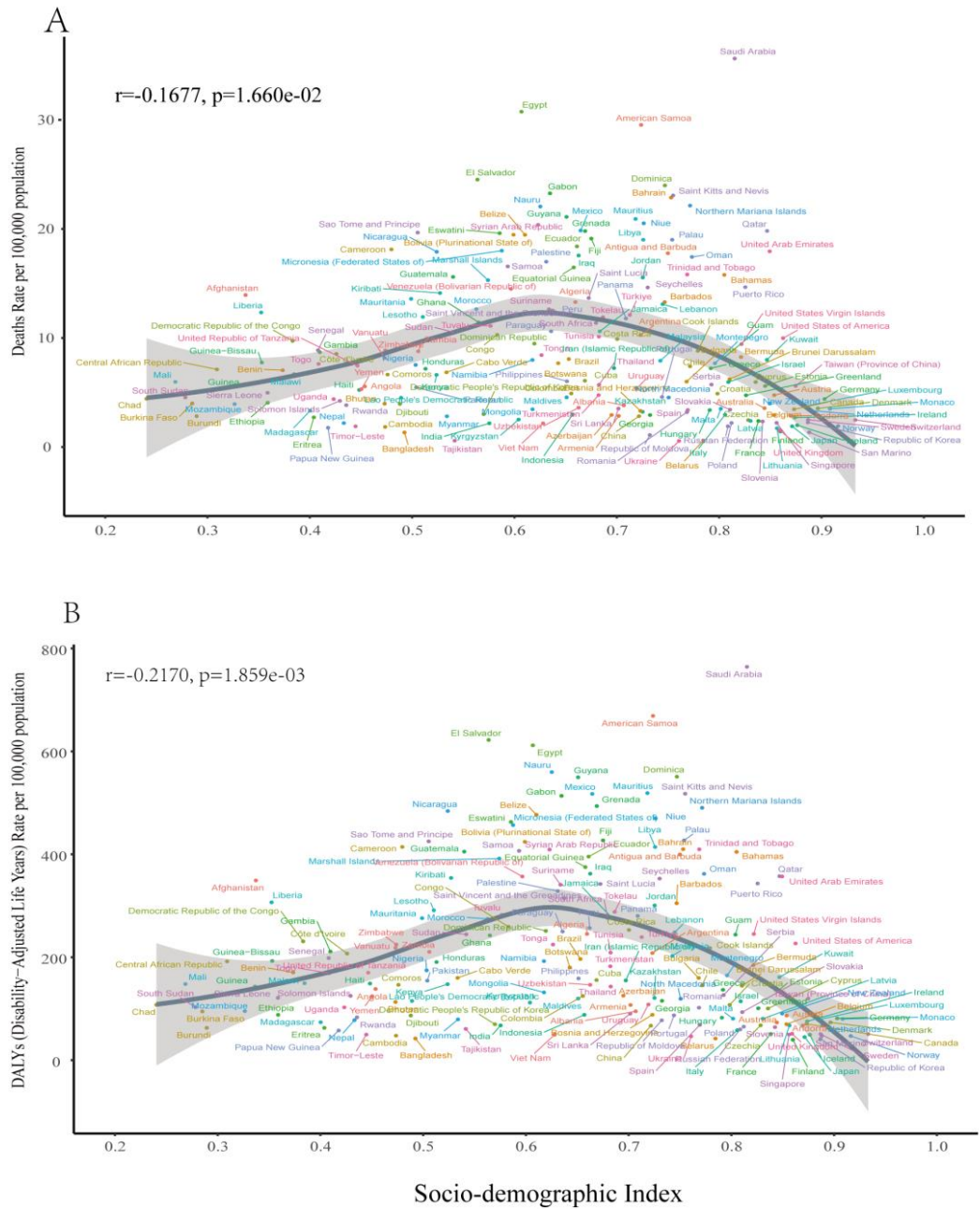

Figure .2. Global CKD burden attributable to high BMI across 204 countries and territories by the SDI in 2021.Global CKD burden attributable to high BMI across 204 countries and territories by the SDI for both sexes combined in 2021.(A) Age-standardized mortality rate (ASMR); and (B) Age-standardized disability-adjusted life year rate (ASDR).

**CKD burden attributable to high BMI by SDI regions**

Compared to regions across the five SDI levels, the ASMR

(Age-Standardized Mortality Rate) was highest in the Middle SDI region and lowest in the High-middle SDI region in 2021 (Figure 4A). From 1990 to 2021, the ASMR increased globally and across all five SDI regions, with the largest increase observed in the Low-middle SDI region (105.2%). During the same period, the absolute number of CKD (Chronic Kidney Disease) deaths attributable to high BMI significantly increased in all SDI regions.

In 2021, the ASDR (Age-Standardized DALY Rate) was highest in the Low-middle SDI region and lowest in the High SDI region (Figure 4B). From 1990 to 2021, the ASDR increased substantially across all SDI regions, with the largest increase seen in the Low-middle SDI region (107.8%). Additionally, the absolute number of DALYs (Disability-Adjusted Life Years) increased in all SDI regions between 1990 and 2021. In areas with relatively low SDI, ASMR and ASDR are high because of insufficient medical resources and limited health services [24], which makes early CKD undetectable in time and unable to be treated regularly after diagnosis.

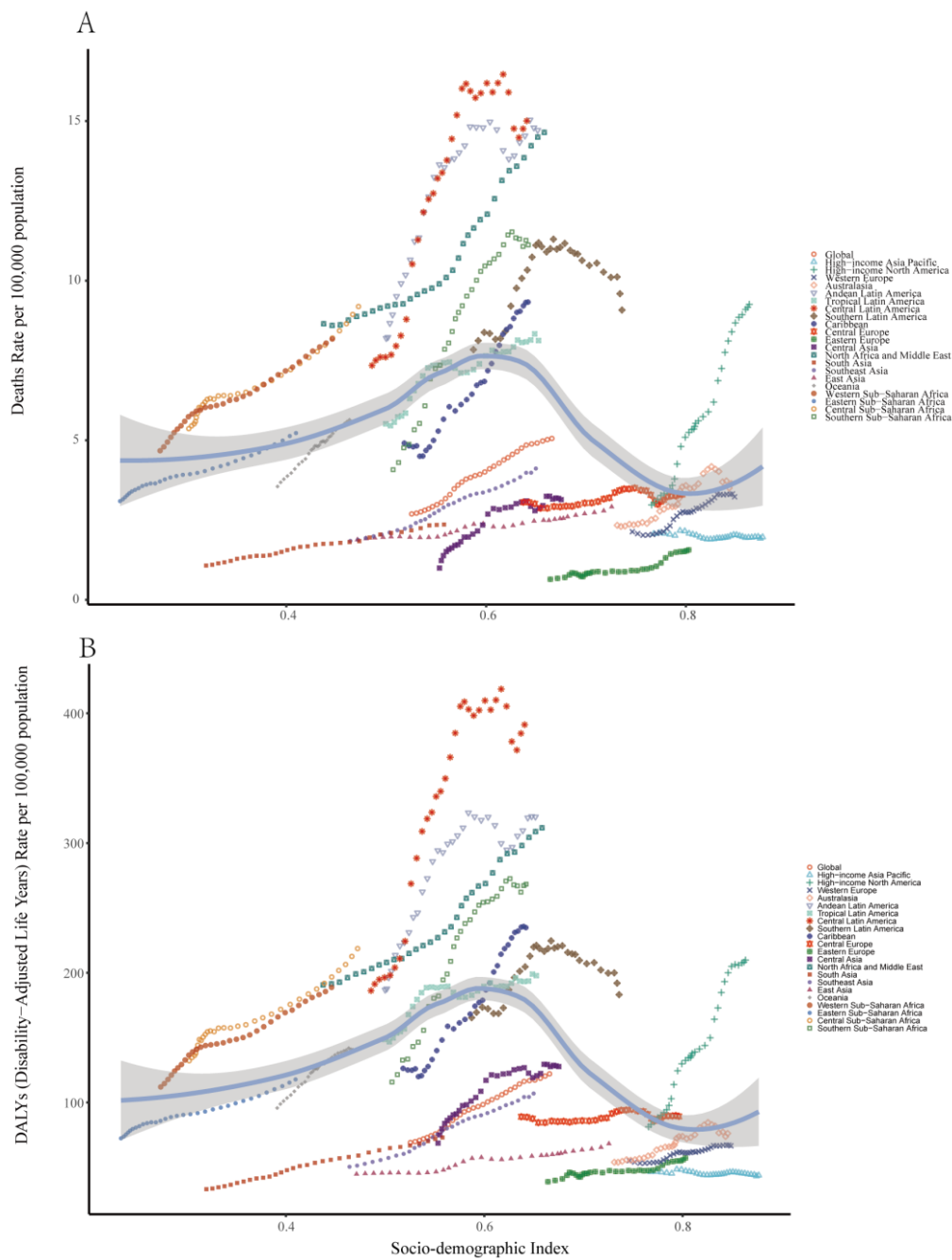

FIGURE .3. The ASMR and ASDR attributable to high BMI across 21 GBD regions by the SDI .Age-standardized mortality rate (ASMR) and age-standardized disability-adjusted life year rate (ASDR) attributable to high body mass index across 21 GBD regions by the socio-demographic index for both sexes combined, 1990–2021. (A) ASMR; and (B) ASDR.

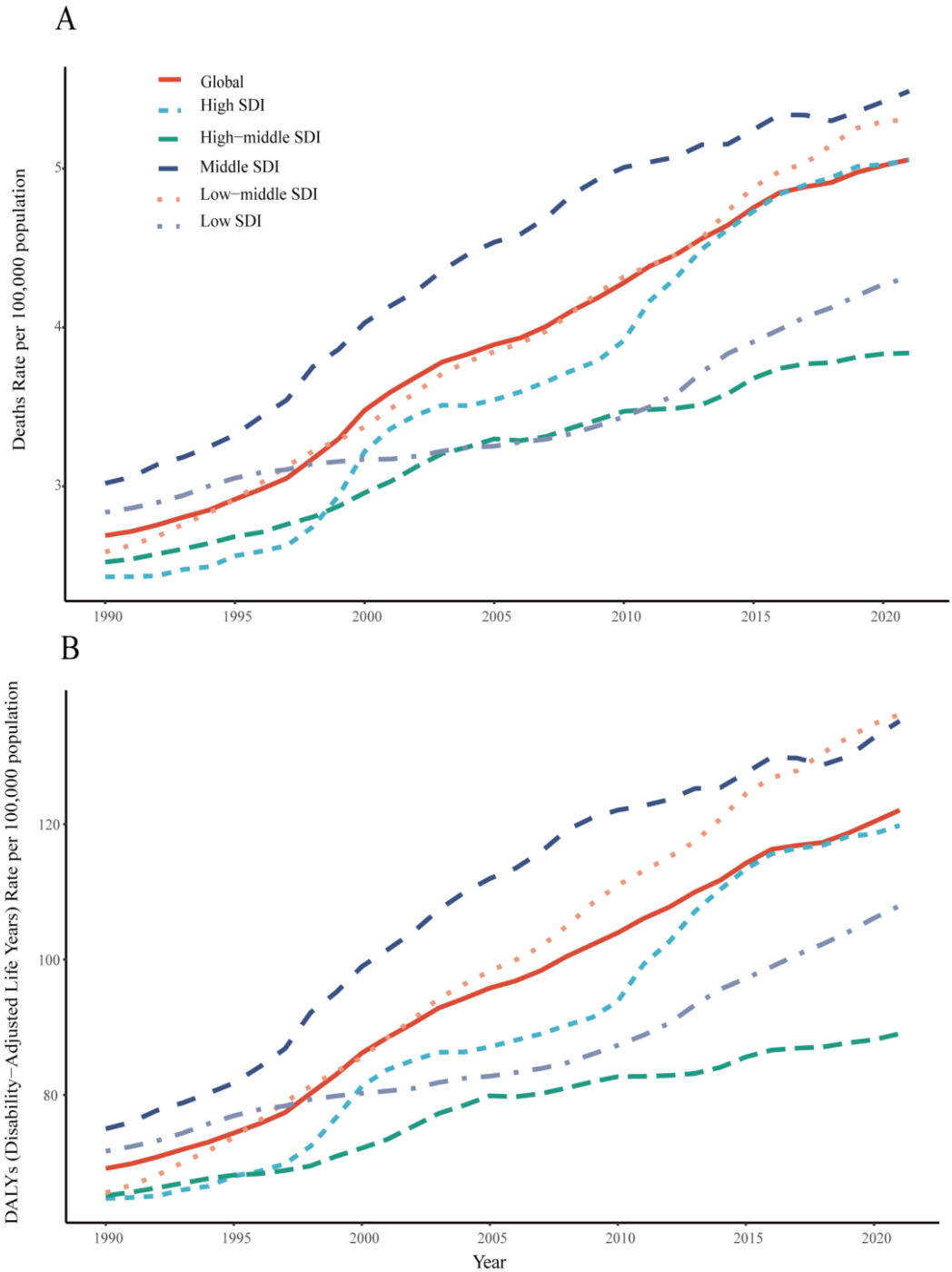

FIGURE .4. The changes of ASMR and ASDR globally and different SDI regions from 1990 to 2021.
Changes in the age-standardized mortality rate (ASMR) and age-standardized disability-adjusted life year rate (ASDR) of chronic kidney dieases attributable to high body mass index globally and in different socio-demographic index regions from 1990 to 2021. (A) ASMR; and (B) ASDR.

**CKD burden attributable to high BMI by ages and genders**

In 2021, globally, the mortality rate of CKD (Chronic Kidney Disease) associated with high BMI increased with age within the 20-54 age range. The number of CKD deaths among females peaked in the 50-54 age group, while for males, it also peaked in the 50-54 age group. Additionally, the number of DALYs (Disability-Adjusted Life Years) peaked at 50-54 years for both females and males (Figure 5). The number of cases, as well as the ASMR (Age-Standardized Mortality Rate) and ASDR (Age-Standardized DALY Rate), were lower among females than males in the 20-54 age group. Globally, the proportion of CKD deaths and DALYs attributable to high BMI increased with age within the 20-54 age group, reaching a peak in the 50-54 age group (41.8% and 34.8%, respectively). Among these, the High-income Asia Pacific region had the highest proportion of CKD deaths (58.7%) and DALYs (48.5%) attributable to high BMI in the 50-54 age group (Supplementary Figure S1).

Furthermore, globally, the ASMR for males was higher than that for females in both 1990 (2.96 vs. 0.48 per 100,000 population) and 2021 (5.47 vs. 1.10 per 100,000 population) (Supplementary Table S4). Similarly, the ASDR for males was higher than that for females in both 1990 (72.45 vs. 28.90 per 100,000 population) and 2021 (128.58 vs. 63.02 per 100,000 population) (Supplementary Table S5).

Between 1990 and 2021, the global disease burden of CKD attributable to high BMI increased significantly for both males and females, with a more pronounced increase observed among males.

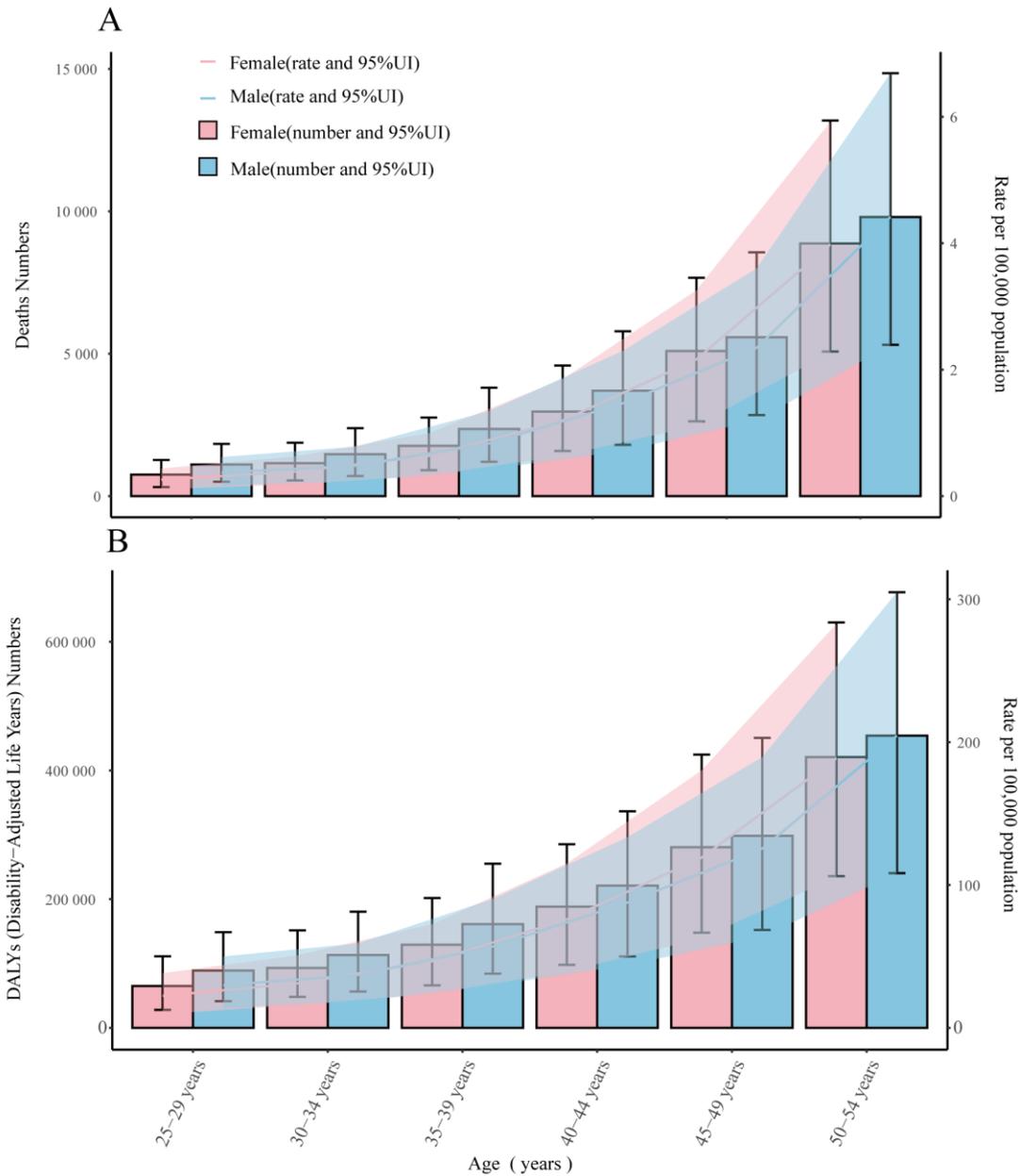

FIGURE .5. Age-specific numbers and rates of deaths and DALYs by age and sex in 2021. Age-specific numbers and rates of deaths and disability-adjusted life years (DALYs) of chronic kidney diseases attributable to high body mass index by age and sex, in 2021. (A) Deaths; and (B) DALYs.

**Decomposition analysis of CKD burden attributable to high BMI**

We conducted a decomposition analysis to further explore the impact of aging, population growth, and epidemiological changes on CKD deaths and DALYs attributable to high BMI. Globally, from 1990 to 2021, the number of deaths increased by 32,622.23. Population growth contributed 12,675.9 (38.86%), aging contributed 3,420.4 (10.48%), and epidemiological changes contributed 16,525.93 (50.66%). In terms of DALYs, the global increase from 1990 to 2021 was 1,792,400.05, with population growth contributing 730,960.12 (40.78%), aging contributing 159,633.93 (8.91%), and epidemiological changes contributing 901,806.03 (50.31%).

The contribution of aging was highest in the Low SDI region, accounting for 455 (14.13%) of deaths, while the contribution of population growth was highest in the Middle SDI region, at 5,459.82 (41%). For DALYs, the contributions of aging and population growth were highest in the Middle SDI region, at 120,602.89 (16.84%) and 278,797.22 (38.94%), respectively. Epidemiological changes had the highest contribution in the Middle SDI and Low-middle SDI regions, with deaths attributable to these changes being 7,695.3 (57.78%) and 5,902.85 (62.52%), respectively. Consistent with their impact on DALYs, the regions with the highest contributions from epidemiological changes were the Middle SDI and Low-middle SDI regions, at 316,593.05 (44.22%) and 261,566.03 (50.58%), respectively (Figure 6).

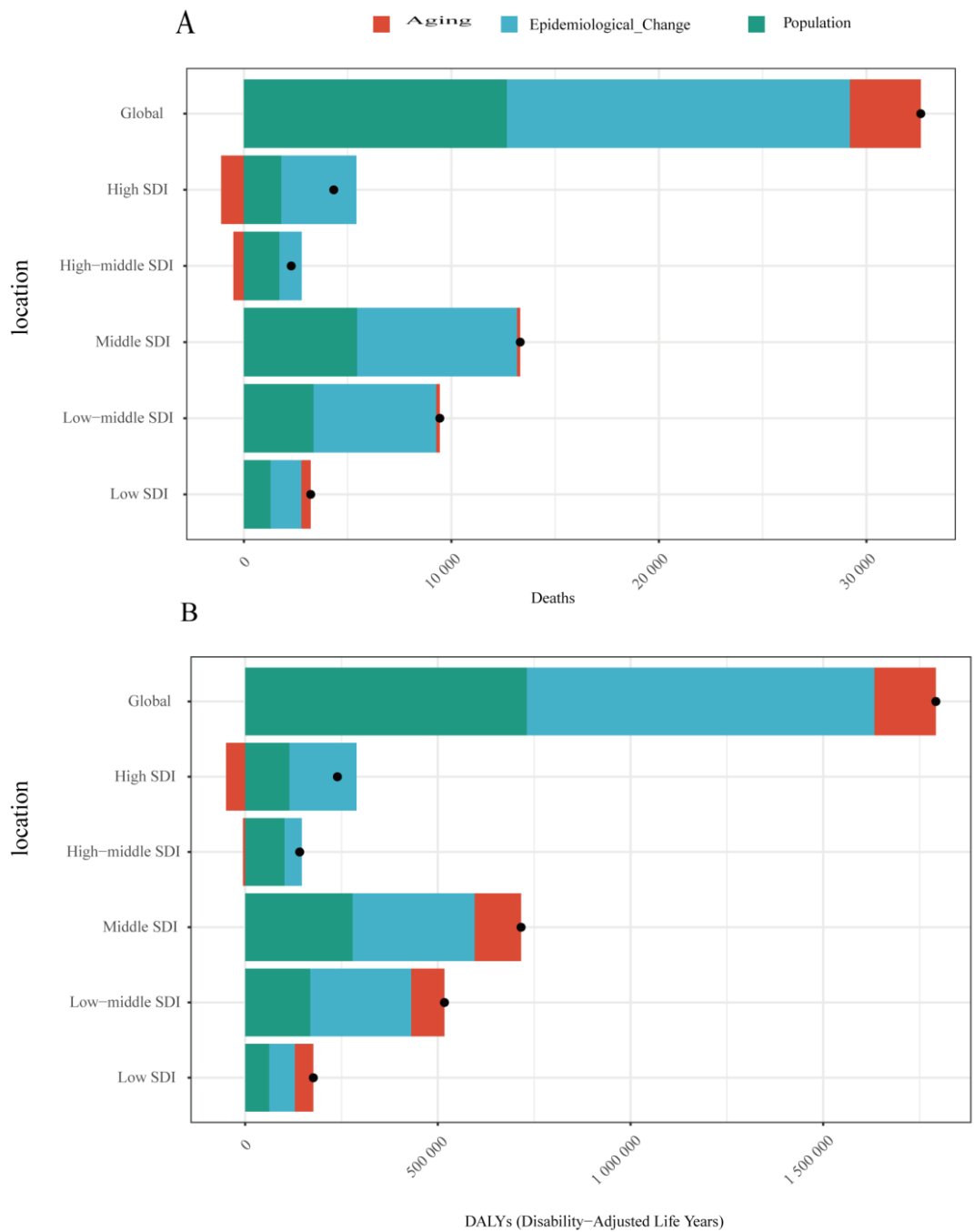

FIGURE .6. Decomposition analysis of changes in the deaths and DALYs between 1990 and 2021.Decomposition analysis of changes in the deaths and DALYs of CKD attributable to high BMI between 1990 and 2021 across SDI regions. DALYs, disability-adjusted life years; CKD, chronic kidney disease; SDI, socio-demographic index.

**Cross-country inequality analysis of CKD burden attributable to high BMI**

There are significant relative and absolute inequalities in the burden of CKD

attributable to high BMI, which are associated with the Socio-demographic Index (SDI). From 1990 to 2021, the Slope Index of Inequality (SII) showed an increasing trend. Regions with lower SDI were observed to have a higher burden of CKD attributable to high BMI. In 1990, the SII was -2.66 (-33.68, 28.37), and by 2021, it increased to -32.28 (-91.68, 27.13), indicating a rise in inequality in disease burden between high SDI and low SDI regions (Figure 7). Additionally, the Concentration Index (CI) was -0.03 (-0.07, 0.02) in 1990 and increased to -0.05 (-0.11, 0.01) in 2021, suggesting a slight rise in the concentration of disease burden, though inequality persists (Figure 7). The results show that most regions exhibited worsening inequalities in SII and CI for countries with lower SDI.

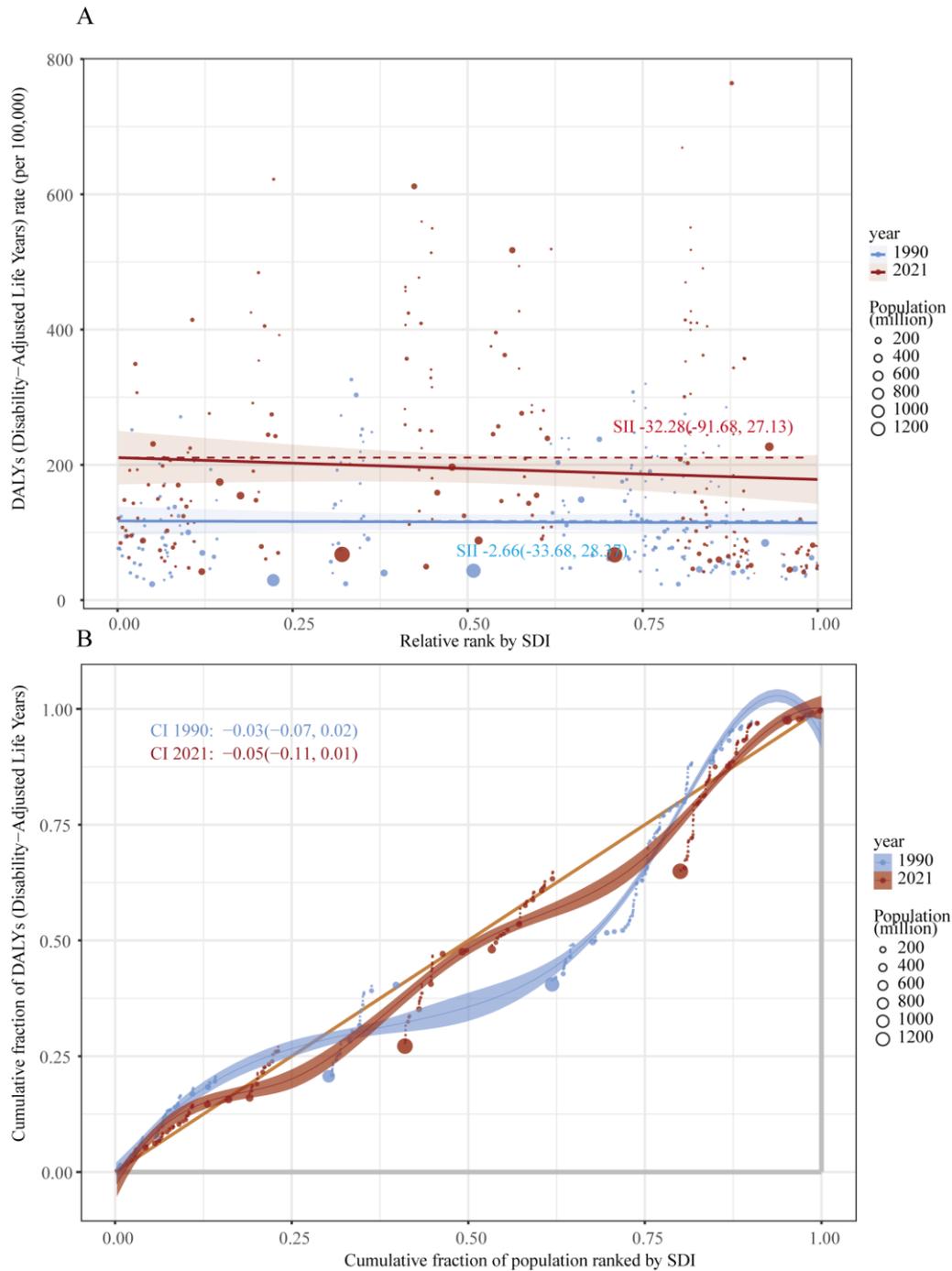

FIGURE .7. Inequality analysis in 1990 and 2021 across the world. Inequality analysis of DALYs in CKD attributable to high BMI in 1990 and 2021 across the world. DALYs, disability-adjusted life years; CKD, chronic kidney disease; SII, slope index of inequality; CI, concentration index.

**Frontier analysis of CKD burden attributable to high BMI**

A frontier analysis was conducted to assess the potential development space for the burden of CKD attributable to high BMI across regions with different SDI levels. Taking into account the SDI of each country, this study identified the top 15 countries or regions with the most development potential, with effective difference (ef_DF) values ranging from 454.02 to 741.45. These countries include Saudi Arabia (741.45), American Samoa (646.09), El Salvador (599.42), Egypt (588.94), Nauru (536.95), Dominica (528.04), Guyana (526.83), Mauritius (496.12), Mexico (494.54), Saint Kitts and Nevis (494.9), Gabon (490.77), Grenada (470.94), Northern Mariana Islands (467.62), Nicaragua (461.29), and Belize (454.02). Low SDI (<0.5) countries located on the frontier include Niger (ef_DF 18.57), Papua New Guinea (ef_DF 34.88), Timor-Leste (ef_DF 26.84), Cambodia (ef_DF 24.8), and Bangladesh (ef_DF 19.3). Additionally, high SDI (>0.85) countries with significant potential for improvement relative to their development stage are the United States of America (ef_DF 203.94), Taiwan (Province of China) (ef_DF 96.76), Austria (ef_DF 63.97), Germany (ef_DF 58.4), and Monaco (ef_DF 57.16) (Figure 8).

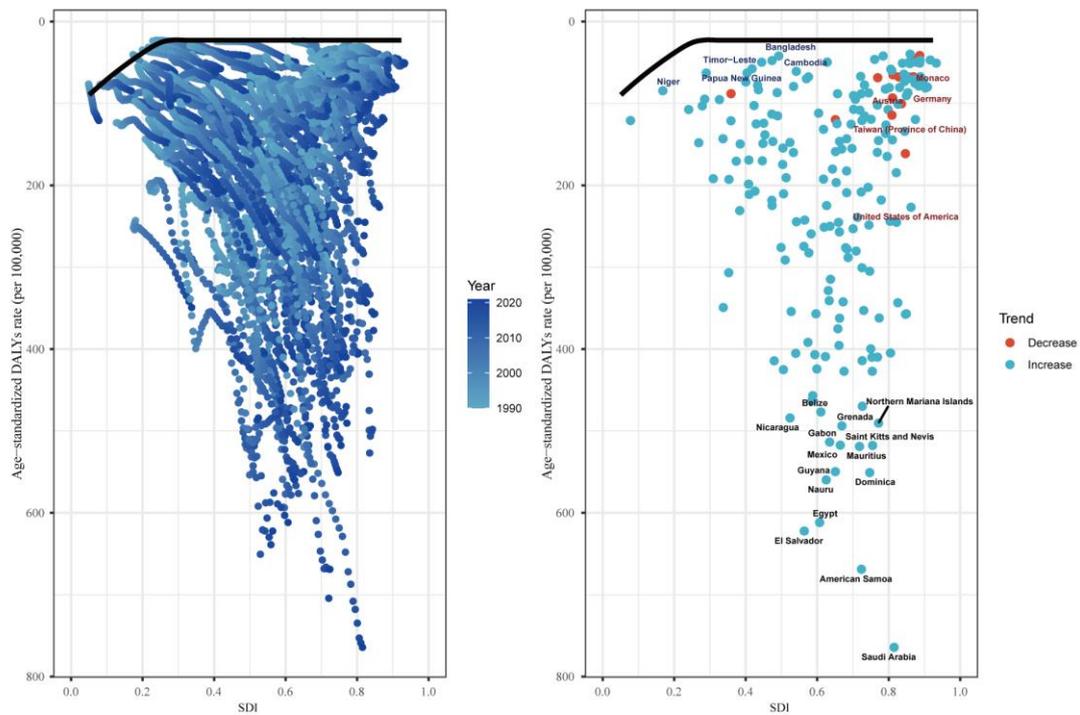

FIGURE .8. Frontier analysis of ASDR based on SDI from 1990 to 2021, and specifically in 2021.Frontier analysis of ASDR in CKD attributable to high BMI based on SDI from 1990 to 2021, and specifically in 2021. ASDR, age-standardized DALY rate; SDI, socio-demographic index.

**Prediction analysis of CKD burden attributable to high BMI**

The BAPC (Bayesian Age-Period-Cohort) model analysis aims to predict the future burden of CKD attributable to high BMI in the absence of interventions. As shown in Figure 9, the global age-standardized mortality rate (ASMR) for CKD attributable to high BMI among individuals aged 20-54 is projected to continue rising over the next decade. By 2035, the global ASMR for both males and females is expected to reach approximately 2.08 (95% CI, 1.88-2.28). The ASMR for males is predicted to increase, reaching 2.27 (95% CI, 2.04-2.49) by 2035. In contrast, the ASMR for females related to high BMI is also expected to rise, with a projected value of 1.88 (95% CI, 1.68-2.08) by 2035, although the overall level for females remains lower than that for males.

    Additionally, the study forecasts age-specific death cases and mortality rates for both males and females. From 2022 to 2035, the number of deaths is expected to increase across all age groups for both genders. For both females and males, mortality rates in the 20-54 age group are projected to show an upward trend.

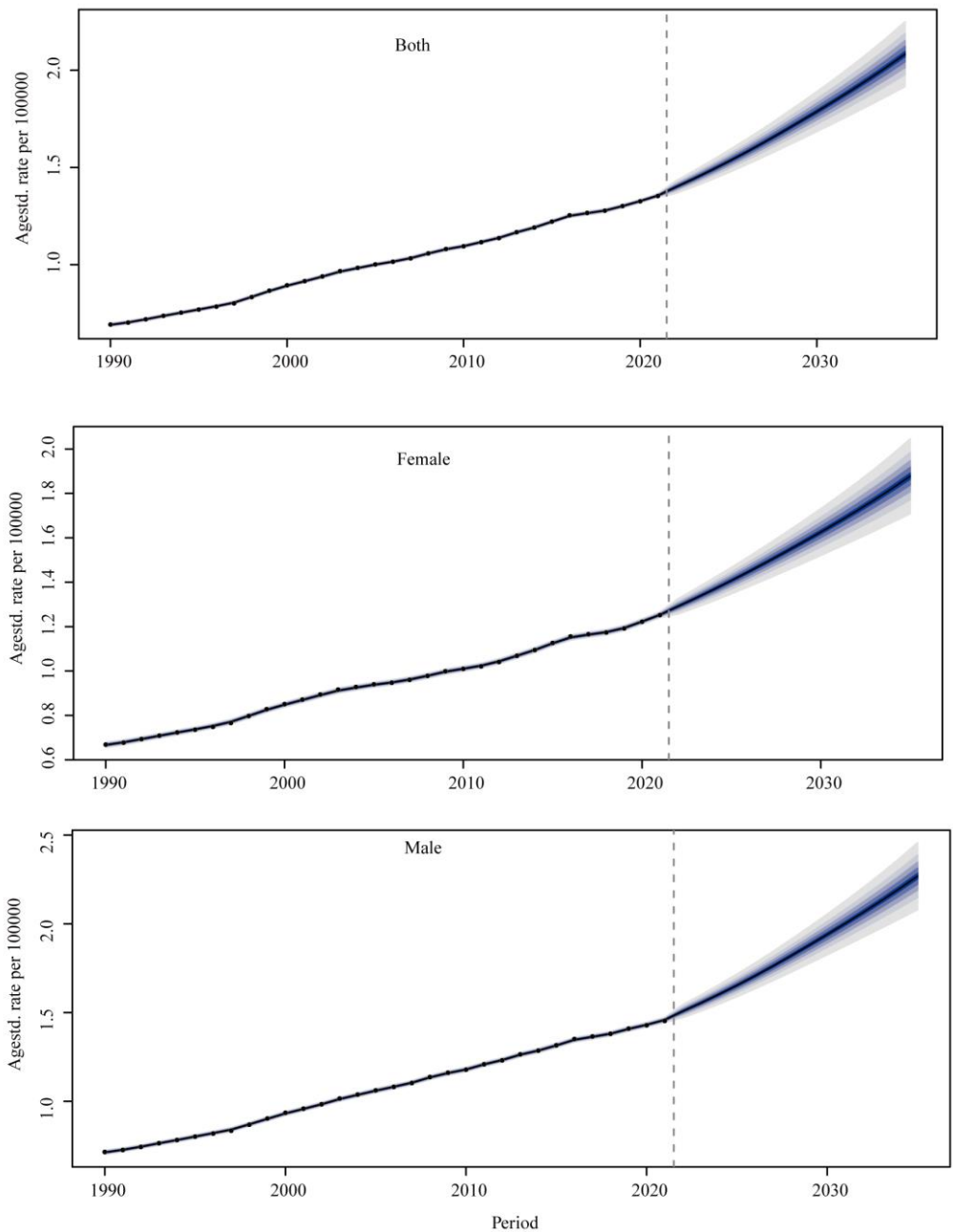

FIGURE .9.Temporal trends of ASMR in GBTCs attributable to high BMI at the global level from 1990 to 2035 for both genders, males, and females. ASMR, age-standardized mortality rate.

## 4. DISCUSSION

This study integrates data from 204 countries and regions to assess global patterns and trends in the burden of chronic kidney disease (CKD) attributable to high BMI over the past 30 years. In 2021, the global number of CKD deaths attributable to high BMI was 44,643.41, and the number of disability-adjusted life years (DALYs) was 2,514,227.16, more than triple the figures from 1990. Additionally, from 1990 to 2021, the global age-standardized mortality rate (ASMR) and age-standardized DALY rate (ASDR) increased significantly by 88.1% and 76.6%, respectively. Among the 20-54 age group examined in this study, males were more susceptible than females, and the global CKD burden attributable to high BMI increased more markedly among males. Within the age groups studied, both the number of deaths and age-standardized rates were higher for males than females. As mentioned in the literature review, the global burden of CKD due to high BMI has been briefly described [25]. However, this study focuses more comprehensively on high BMI as a primary risk factor for disease. Therefore, timely measures must be taken to address high BMI in populations. Since 1980, the global prevalence of obesity has more than doubled. In 2015, approximately 12% of the global adult population (641 million people) was classified as obese. If current trends continue, by 2025, around 18% of men and 21% of women worldwide will be obese[26]. This alarming growth rate underscores the urgent need for effective preventive measures and interventions to address this global public health challenge. Concurrent with the rising obesity rates, the prevalence of CKD has also increased. In 2017, CKD affected over 800 million people globally and remains a leading cause of morbidity and mortality [27].

Previous studies have demonstrated a causal relationship between obesity and CKD [28]. Another large meta-analysis found that elevated BMI, waist circumference, and waist-to-height ratio are independent risk factors associated with kidney function loss, estimated glomerular filtration rate, and individual mortality [29]. Recent research has identified obesity as an independent risk factor for CKD, regardless of the presence of diabetes, arterial hypertension, or other known risk factors. In the United States, nearly half of the population suffers from both CKD and obesity[30]. A meta-analysis by Wang et al. found that obese individuals account for 33% of all kidney disease cases in the U.S., and the analysis predicted that approximately 14% of men and 25% of women in industrialized countries develop CKD due to being overweight or obese [31]. Therefore, there is an urgent need to understand the burden of CKD caused by high BMI. This study provides a comprehensive summary of the global, regional, and national burden of CKD attributable to high BMI, using the latest GBD 2021 data to project trends over the next decade in the absence of interventions. This analysis aims to provide policymakers with resources for effective allocation and targeted prevention strategies.

The increasing trends in overweight and obesity cannot be ignored[32, 33]. Generally, lifestyle and dietary habits are the most significant contributors to obesity, particularly in cases of insufficient daily physical activity and poor dietary patterns, such as high consumption of calorie-dense foods, carbohydrate-rich soft drinks, and fast food, coupled with inadequate intake of fresh fruits and vegetables[34]. In our study, Saudi Arabia had the highest ASMR and ASDR for CKD attributable to high BMI in 2021. Over the past few decades, the prevalence of obesity in Saudi Arabia has been rising, driven by several factors. On one hand, approximately 20% of Saudis are obese, with the majority being older married women[35]. On the other hand, urbanization and economic development have led to lifestyle changes, including increased sedentary behavior and a shift toward Western dietary habits high in fat, salt, and sugar [36, 37]. In contrast, some low-development countries face social, cultural, economic, political, and environmental challenges that result in a lack of knowledge and understanding of healthy diets, limited access to high-quality healthy foods, and frequent consumption of processed foods, leading to high obesity rates and elevated BMI levels. In comparison, Ukraine had the lowest ASMR, and Finland had the lowest ASDR. Finland's diet is aligned with the Nordic dietary pattern, which emphasizes environmental protection and sustainability, with plant-based proteins, fish, and seafood as primary food sources[38]. Additionally, Nordic countries jointly published the Nordic Nutrition Recommendations (NNR), providing dietary guidelines and nutrient intake reference values. Therefore, more policies are needed at both the population and national leadership levels to reduce the impact of high BMI on CKD.

At the GBD regional level, over the past 30 years, only High-income Asia Pacific saw a slight decline in ASMR for CKD attributable to high BMI, while Eastern Europe experienced a slight decline in ASDR. All other 21 GBD regions showed significant increases in both ASMR and ASDR. As the results indicate, the regions with the highest disease burden were Andean Latin America, Central Latin America, and North Africa and the Middle East. Lower levels of social development and healthcare may exacerbate the burden of BMI-related CKD due to insufficient clinical laboratory diagnostics, shortages of medical personnel, limited access to medications, and the lack of universal health insurance [39]. The regions with the lowest ASMR and ASDR were Eastern Europe, followed by High-income Asia Pacific.

The Socio-demographic Index (SDI) significantly influences the geographic distribution of CKD burden related to high BMI. This study reveals that from 1990 to 2021, the inequality in DALYs between high SDI and low SDI regions became more pronounced. This phenomenon is partly attributed to countries with higher SDI having more advanced diagnostic facilities and comprehensive disease screening mechanisms, while low SDI countries face relative shortages of medical resources and inadequate healthcare systems.

Furthermore, age and gender analyses revealed distinct outcomes. In the 20-54 age group, both ASMR and ASDR for CKD attributable to high BMI increased with age, and males across all age groups exhibited higher ASMR and ASDR. The future projection analysis in this study indicates that, in the absence of interventions, ASMR for both females and males will rise over the next decade, but male mortality rates are expected to remain higher across all age groups. Therefore, additional medical resources must be allocated to ensure timely and effective treatment for males, along with proactive measures to control obesity rates, thereby reducing the burden of CKD related to high BMI in this population.

Frontier analysis highlights significant disparities among countries in managing the burden of CKD attributable to high BMI. Notably, low SDI countries such as Niger, Papua New Guinea, Timor-Leste, Cambodia, and Bangladesh have achieved remarkable results in managing the burden. In contrast, high SDI countries like the United States of America, Taiwan (Province of China), Austria, Germany, and Monaco have not met expectations in addressing CKD burden related to high BMI. Geographic factors and dietary habits may contribute to this inconsistency. This situation underscores the urgent need for these high SDI countries to optimize and reform health policy formulation and implementation more effectively.

Despite covering 204 countries and regions, this study has several limitations. First, due to the use of a large number of data sources of varying quality, GBD estimates of CKD burden attributable to high BMI may deviate to some extent from actual data. Second, the assessment of CKD burden did not account for comorbidities of CKD, and this study did not investigate CKD caused by different etiologies. Third, some countries may lack sufficient data, leading to the use of GBD modeling to estimate CKD burden attributable to high BMI in these regions.

Conclusion

In summary, high BMI is a significant factor contributing to the burden of CKD. From 1990 to 2021, both the standardized and absolute values of CKD burden attributable to high BMI increased substantially worldwide, particularly among males, in High-income North America, and in Low-middle SDI regions. Our findings will better guide relevant government departments in formulating appropriate measures to manage CKD and high BMI in different regions in the future, thereby improving public health issues.

## AUTHOR CONTRIBUTIONS

Yu Chen: Data curation (equal); formal analysis (equal); investigation (equal); project administration (equal); writing – original draft (lead). Guangxi Wu: Data curation (equal); formal analysis (equal); project administration (equal); writing – original draft (equal). Wei Ma: Data curation (equal); project administration (equal). Ying Zhang: Data curation (equal); formal analysis (equal); investigation (equal); project administration (equal); writing – review and editing (equal).

## ACKNOWLEDGMENTS

The authors greatly thank to the work by the 2021 GBD collaborators.

## CONFLICT OF INTEREST STATEMENT

The authors declare no competing interests.

## DATA AVAILABILITY STATEMENT

All data used in this article are all publicly available at online GBD repository (http://ghdx.healthdata.org/gbd-results-tool).

## FUNDING INFORMATION

The author(s) declare that no financial support was received for the research, authorship, and/or publication of this article.

## ETHICS STATEMENT

Not applicable.

## REFERENCES

1. Sundström J, Bodegard J, Bollmann A.Prevalence, outcomes, and cost of

chronic kidney disease in a contemporary population of 2·4 million patients from 11 countries: The CaReMe CKD study.The Lancet Regional Health - Europe 2022, 20(30):100438.

2. Abbas KM, Aboyans V, Ackerman IN, Adair T, Adebayo OM, Adelson JD, Afshin A, Ahmadi A, Alahdab F, Alipour V,et al.Global burden of 369 diseases and injuries in 204 countries and territories, 1990-2019: a systematic analysis for the Global Burden of Disease Study 2019.LANCET 2020, 396(10258):1204-1222.

3. Foreman KJ, Marquez N, Dolgert A, Fukutaki K, Fullman N, McGaughey M, Pletcher MA, Smith AE, Tang K, Yuan CW,et al.Forecasting life expectancy, years of life lost, and all-cause and cause-specific   mortality for 250 causes of death: reference and alternative scenarios for   2016-40 for 195 countries and territories.LANCET 2018, 392(10159):2052-2090.

4. Liyanage T, Ninomiya T, Jha V, Neal B, Patrice HM, Okpechi I, Zhao MH, Lv J, Garg AX, Knight J,et al.Worldwide access to treatment for end-stage kidney disease: a systematic review.LANCET 2015, 385(9981):1975-1982.

5. Brown EA, Zhao J, McCullough K, Fuller DS, Figueiredo AE, Bieber B, Finkelstein FO, Shen J, Kanjanabuch T, Kawanishi H,et al.Burden of Kidney Disease, Health-Related Quality of Life, and Employment Among   Patients Receiving Peritoneal Dialysis and In-Center Hemodialysis: Findings From   the DOPPS Program.AM J KIDNEY DIS 2021, 78(4):489-500.

6. Saran R, Pearson A, Tilea A, Shahinian V, Bragg-Gresham J, Heung M, Hutton DW, Steffick D, Zheng K, Morgenstern H,et al.Burden and Cost of Caring for US Veterans With CKD: Initial Findings From the VA   Renal Information System (VA-REINS).AM J KIDNEY DIS 2021, 77(3):397-405.

7. Gadde KM, Martin CK, Berthoud HR, Heymsfield SB.Obesity: Pathophysiology and Management.J AM COLL CARDIOL 2018, 71(1):69-84.

8. Dai H, Alsalhe TA, Chalghaf N, Ricco M, Bragazzi NL, Wu J.The global burden of disease attributable to high body mass index in 195   countries and territories, 1990-2017: An analysis of the Global Burden of Disease Study.PLOS MED 2020, 17(7):e1003198.

9. Tsai AG, Williamson DF, Glick HA.Direct medical cost of overweight and obesity in the USA: a quantitative   systematic review.OBES REV 2011, 12(1):50-61.

10. Singh-Manoux A, Fayosse A, Sabia S, Tabak A, Shipley M, Dugravot A, Kivimaki M.Clinical, socioeconomic, and behavioural factors at age 50 years and risk of   cardiometabolic multimorbidity and mortality: A cohort study.PLOS MED 2018, 15(5):e1002571.

11. Freisling H, Viallon V, Lennon H, Bagnardi V, Ricci C, Butterworth AS, Sweeting M, Muller D, Romieu I, Bazelle P,et al.Lifestyle factors and risk of multimorbidity of cancer and cardiometabolic   diseases: a multinational cohort study.BMC MED 2020, 18(1):5.

12. Zhou XD, Targher G, Byrne CD, Somers V, Kim SU, Chahal C, Wong VW, Cai J, Shapiro MD, Eslam M,et al.An international multidisciplinary consensus statement on MAFLD and the risk of   CVD.HEPATOL INT 2023, 17(4):773-791.
13. Hendren NS, de Lemos JA, Ayers C, Das SR, Rao A, Carter S, Rosenblatt A, Walchok J, Omar W, Khera R,et al.Association of Body Mass Index and Age With Morbidity and Mortality in Patients   Hospitalized With COVID-19: Results From the American Heart Association COVID-19   Cardiovascular Disease Registry.CIRCULATION 2021, 143(2):135-144.
14. Alsaqaaby MS, Cooney S, le Roux CW, Pournaras DJ.Sex, race, and BMI in clinical trials of medications for obesity over the past   three decades: a systematic review.LANCET DIABETES ENDO 2024, 12(6):414-421.
15. Murray C.Findings from the Global Burden of Disease Study 2021.LANCET 2024, 403(10440):2259-2262.
16. Fitzmaurice C, Allen C, Barber RM, Barregard L, Bhutta ZA, Brenner H, Dicker DJ, Chimed-Orchir O, Dandona R, Dandona L,et al.Global, Regional, and National Cancer Incidence, Mortality, Years of Life Lost,   Years Lived With Disability, and Disability-Adjusted Life-years for 32 Cancer   Groups, 1990 to 2015: A Systematic Analysis for the Global Burden of Disease   Study.JAMA ONCOL 2017, 3(4):524-548.
17. Vos T, Allen C, Arora M, Barber RM, Bhutta ZA, Brown A, Carter A, Casey DC, Charlson FJ, Chen AZ,et al.Global, regional, and national incidence, prevalence, and years lived with disability for 310 diseases and injuries, 1990–2015: a systematic analysis for the Global Burden of Disease Study 2015.The Lancet 2016, 388(10053):1545-1602.
18. Zhao S, Wang H, Chen H, Wang S, Ma J, Zhang D, Shen L, Yang X, Chen Y.Global magnitude and long-term trend of ischemic heart disease burden attributed   to household air pollution from solid fuels in 204 countries and territories,   1990-2019.INDOOR AIR 2022, 32(2):e12981.
19. Collaborators GDAI.Global, regional, and national incidence, prevalence, and years lived with   disability for 354 diseases and injuries for 195 countries and territories,   1990-2017: a systematic analysis for the Global Burden of Disease Study 2017.LANCET 2018, 392(10159):1789-1858.
20. Cole TJ, Lobstein T.Extended international (IOTF) body mass index cut-offs for thinness, overweight   and obesity.PEDIATR OBES 2012, 7(4):284-294.
21. Hung GY, Horng JL, Yen HJ, Lee CY, Lin LY.Changing incidence patterns of hepatocellular carcinoma among age groups in   Taiwan.J HEPATOL 2015, 63(6):1390-1396.
22. Liu Z, Jiang Y, Yuan H, Fang Q, Cai N, Suo C, Jin L, Zhang T, Chen X.The trends in incidence of primary liver cancer caused by specific etiologies: Results from the Global Burden of Disease Study 2016 and implications for liver   cancer prevention.J HEPATOL 2019, 70(4):674-683.
23. Chevan A, Sutherland M.Revisiting Das Gupta: Refinement and Extension of Standardization and Decomposition.DEMOGRAPHY 2009, 46(3):429-449.
24. Ejemot-Nwadiaro RI, Ehiri JE, Arikpo D, Meremikwu MM, Critchley JA.Hand-washing promotion for preventing diarrhoea.COCHRANE DB SYST

REV 2021, 12(1):CD4265.
25. Yau K, Kuah R, Cherney D, Lam T.Obesity and the kidney: mechanistic links and therapeutic advances.NAT REV ENDOCRINOL 2024, 20(6):321-335.
26. Collaborators GDAI.Worldwide trends in body-mass index, underweight, overweight, and obesity from   1975 to 2016: a pooled analysis of 2416 population-based measurement studies in   128.9 million children, adolescents, and adults.LANCET 2017, 390(10113):2627-2642.
27. Kovesdy CP.Epidemiology of chronic kidney disease: an update 2022.KIDNEY INT SUPPL 2022, 12(1):7-11.
28. Ye C, Kong L, Zhao Z, Li M, Wang S, Lin H, Xu Y, Lu J, Chen Y, Xu Y,et al.Causal Associations of Obesity With Chronic Kidney Disease and Arterial Stiffness: A Mendelian Randomization Study.J CLIN ENDOCR METAB 2022, 107(2):e825-e835.
29. Chang AR, Grams ME, Ballew SH, Bilo H, Correa A, Evans M, Gutierrez OM, Hosseinpanah F, Iseki K, Kenealy T,et al.Adiposity and risk of decline in glomerular filtration rate: meta-analysis of   individual participant data in a global consortium.BMJ-BRIT MED J 2019, 364(10):k5301.
30. Kreiner FF, Schytz PA, Heerspink H, von Scholten BJ, Idorn T.Obesity-Related Kidney Disease: Current Understanding and Future Perspectives.BIOMEDICINES 2023, 11(9):9-2498.
31. Wang Y, Chen X, Song Y, Caballero B, Cheskin LJ.Association between obesity and kidney disease: a systematic review and   meta-analysis.KIDNEY INT 2008, 73(1):19-33.
32. Bulik CM, Hardaway JA.Turning the tide on obesity?SCIENCE 2023, 381(6657):463.
33.Lingvay I, Cohen RV, Roux C, Sumithran P.Obesity in adults.LANCET 2024, 404(10456):972-987.
34. Sinha A, Kling S.A review of adolescent obesity: prevalence, etiology, and treatment.OBES SURG 2009, 19(1):113-120.
35. Salem V, AlHusseini N, Abdul RH, Naoum A, Sims OT, Alqahtani SA.Prevalence, risk factors, and interventions for obesity in Saudi Arabia: A systematic review.OBES REV 2022, 23(7):e13448.
36. Althumiri NA, Basyouni MH, AlMousa N, AlJuwaysim MF, Almubark RA, BinDhim NF, Alkhamaali Z, Alqahtani SA.Obesity in Saudi Arabia in 2020: Prevalence, Distribution, and Its Current   Association with Various Health Conditions.HEALTHCARE-BASEL 2021, 9(3):3-311.
37. Alhusseini N, Alsinan N, Almutahhar S, Khader M, Tamimi R, Elsarrag MI, Warar R, Alnasser S, Ramadan M, Omair A,et al.Dietary trends and obesity in Saudi Arabia.FRONT PUBLIC HEALTH 2023, 11(11):1326418.
38. Mithril C, Dragsted LO, Meyer C, Tetens I, Biltoft-Jensen A, Astrup A.Dietary composition and nutrient content of the New Nordic Diet.PUBLIC HEALTH NUTR 2013, 16(5):777-785.
39. Wilson ML, Fleming KA, Kuti MA, Looi LM, Lago N, Ru K.Access to pathology and laboratory medicine services: a crucial gap.LANCET 2018,

391(10133):1927-1938.

Declaration of Interest form

The authors declare no competing interests.